\documentclass[11pt]{article} 

\usepackage[psamsfonts]{amssymb}

\newcommand{\sig}{\sigma}

\newcommand{\diff}{\partial}
\newcommand{\ba}{\begin{array}}
\newcommand{\ea}{\end{array}}

\newcommand{\bs}[1]{{\boldsymbol{#1}}}

\newcommand{\Bdot}{{\,\bs{\cdot}\,}}

\newcommand{\Bk}{{\bs{k}}}

\newcommand{\Bm}{{\bs{m}}}

\newcommand{\Bn}{{\bs{n}}}

\newcommand{\Bq}{{\bs{q}}}

\newcommand{\Br}{{\bs{r}}}

\newcommand{\bC}{{\mathbb{C}}}

\newcommand{\bH}{{\mathbb{H}}}

\usepackage{amsbsy}

\newcommand{\beqa}[1]{\begin{eqnarray}\label{#1}}
\newcommand{\eeqa}{\end{eqnarray}}
\newcommand{\be}[1]{\begin{equation}\label{#1}}
\newcommand{\ee}{\end{equation}}

\newcommand{\Eq}[1]{Eq.\,\ref{#1}}
\newcommand{\eq}[1]{(\ref{#1})}

\begin{document} \large
 
\begin{center}  {\bf Is the Statistical
Interpretation of Quantum Mechanics\\ Implied by the Correspondence Principle
?}\end{center}
\normalsize
\begin{center} {\sl Kurt Gottfried\footnote{ Presented at the conference
``Epistemological and Experimental Perspectives on Quantum Physics,'' Vienna, September
3-6, 1998; to appear in 7th. Yearbook, Institute Vienna Circle, D. Greenberger,
W.L. Reiter and A. Zeilinger (eds.), Kluwer: Dordrecht 1999.}
\\Laboratory for Nuclear Studies, Cornell University, Ithaca NY 14853}\\November 29,
1998\end{center}
\vspace{  .5cm}

Our impresarion, Anton Zeilinger, ignored my pleas of ignorance and prevailed
on me to talk about my discussions with John Bell about the foundations of quantum
mechanics.  This `debate' was  aborted by John's tragic death shortly after we last met at
a wonderful workshop in Amherst attended by several people in this audience.  At that
time, John's last paper ``Against Measurement'' was about to be published.\cite{Bell}  It
featured a wonderfully barbed attack on the treatment of measurement in my 1966 textbook. 
I was delighted that the most profound student of quantum mechanics since the Founding
Fathers, and
 an old friend from CERN, had paid close attention to what I had
written, because   with but one exception, \cite{Cini} no one publishing
in  the field had ever mentioned my work even when espousing a position
that I had taken long before.  
 
 John would not have changed his views had he been able to hear my response 
at the CERN memorial symposium.\cite{KG}   Furthermore,  
 I was far from
satisfied with what I published back then.  On the other hand, I continue to find
John's critique of orthodoxy to be rather less overwhelming than  his superb
rhetoric. Nevertheless, what I will say was
stimulated by reflecting on the views John espoused in our last
conversations and in his last paper.

\begin{center}  {\bf I}\end{center}

In large part, the interpretation of quantum mechanics is  so controversial  because
the basic  equations of the theory seem not to say 
what they mean in terms of pre-existing concepts. In {\em retrospect}, the
{\em  successful} developments of physics followed a
clear path from the {\em  Principia} until  it
 disappeared  into a fog in 1925.   
Whatever Thomas Kuhn and his disciples may say, before the advent of quantum mechanics 
physics had ultimately proven to be a cumulative pursuit, with new knowledge built on prior
concepts. 

Newton's equations of motion  defined all its
new concepts in terms of then  accepted conceptions of space and  time.
The next giant steps concerned heat  and electromagnetism. Statistical
mechanics, while full of subtleties, nevertheless demonstrated how thermodynamics
could be related to an underlying Newtonian description. In electrodynamics, the meaning
of the new concept, the electromagnetic field, is defined by Maxwell's equations
and the Lorentz force law by means of Newton's equations. Admitedly, special relativity
introduced a new conception of space and time, but their meaning 
was  defined by a more penetrating examination of   pre-existing concepts using familiar
tools: measuring rods and clocks.   General relativity, while profoundly new, is, if you
pardon the expression, trivial, provided you are freely falling and short-sighted.
Furthermore, Einstein's equations tell you that -- you do not need him whispering in your
ear.\cite{Whig}

Quantum mechanics does not fit this pattern, it would appear. 
On its own, the formalism does not seem to disclose what the wave function means,
and gurus such  as  Born, Bohr and Heisenberg were  needed to translate
 the equations   into our language. I acknowledged this in my response to John:\cite{KG}
\begin{quote} If one were to hand the Schr\"odinger equation to Maxwell, and tell him that
it describes the structure of matter in terms of various point particles whose masses and
charges are to be seen in the equation, this knowledge would not, by itself, enable
Maxwell to figure out what is meant by the wave function. Eventually he would need help:
``Oh, I forgot to tell you that according to Rabbi Born, a great thinker in the
yeshiva that flourished in G\"ottingen in the early part of the 20th century,
$|\Psi(\Br_1,\ldots,t)|^2 $ is \ldots'' \end{quote}

My aim is to explore whether the situation is really this bleak -- {\em to ask to what
extent, if any, the statistical interpretation of quantum mechanics is revealed by the
theory's formalism.}  I make no claim to
having a fully satisfactory answer to this question.   What
 I have to say is really the reading  of an old book
from the back towards the front.    
\vspace{  .5cm}

 I now put myself into the shoes of this fictional Maxwell. The task of
fathoming the physical content of  the Schr\"odinger equation  leads me to  adopt the
following assumptions and ground rules: 
\begin{itemize}  \item The {\em mathematical} formalism of orthodox
quantum mechanics provides a  complete and 
consistent description of Nature as it stands.  The  implications of the
formalism, such as  the superposition principle, are not to be tampered with.
\item   It is not permissible to invoke the  statistical interpretation,  the
collapse of the wave function,   or the influence of some dynamical environment.
\cite{Bell-2}
\item If quantum mechanics is indeed complete, there must exist conditions under which
classical mechanics provides an essentially exact approximation to quantum mechanics for
systems having properties that are defined by this very limit.\cite{FAPP} 
 \end{itemize}
\noindent{\em The goal is to examine the classical limit of the quantum mechanical
formalism to learn the extent to which this limit  compels the statistical interpretation.}
If such a link could be established  it would continue the tradition of
connecting new developments in physics  to the conceptual roots of the
discipline.

I will present an argument which claims that the Schr\"odinger equation, when examined in
the classical limit, leads to the statistical interpretation for degrees of freedom
described by finite-dimensional Hilbert spaces  having no classical counterpart. In
contrast, the argument, as it stands here, does not  lead to the statistical interpretation
for the degrees of freedom that have a classical
counterpart.  

 Incidentally, my title is
purposely close to that of  a remarkable paper by Van Vleck written 70 years ago,  ``The
Correspondence Principle in the Statistical Interpretation of Quantum Mechanics,''
 which had a somewhat similar goal.\cite{VV}  So I am singing for my supper from
an old though unfinished score. 

\newpage

\begin{center}  {\bf II}\end{center}

 Schr\"odinger's very first equation in his first paper on wave mechanics is the  classical
Hamilton-Jacobi equation; then he made the great  leap to his wave equation. Is there a
return path?

The first step is to note that the Schr\"odinger equation implies  that in the
naive classical limit 
$\hbar \to 0$, the wave function $\Psi$ has an essential singularity,\cite{Birkhoff}
which then motivates  the WKB Ansatz
\be {1} \Psi = e^{i\Theta(\Bq,t)/\hbar}\;,\qquad \Theta = S+
\frac{\hbar}{i}U+ O(\hbar^2)\;, 
\ee  where  $\Bq \equiv (\Bq_1,\ldots,\Bq_N)$
is a point in the configuration space $\bC$ of an $N$ particle system. At this first stage,
I  ignore  ``internal'' variables, such as spin, inhabiting  finite
dimensional Hilbert spaces with no classical counterpart.  

\vspace{  .3cm} 

\noindent The  Schr\"odinger's equation then produces the following
familiar {\em  facts}:
\begin{enumerate}
\item Ansatz \eq{1} is only legitimate --  only produces a respectable asymptotic
series, if
\be {4} \frac{|\diff \Theta/\diff \Bq_k|^2}{|\diff ^2 \Theta/\diff \Bq_l^2|} \ll
\hbar\;;\ee
when this condition is violated the  approximation \eq{1} is not valid, and the
Schr\"odinger equation itself must be used. 
\item The leading term  $S$ satisfies the classical Hamilton-Jacobi
equation in the region accessible to classical motion. Recall that the classical
trajectories are the curves in $\bC$ everywhere normal to the surfaces of constant $S$.
Thus
$\Psi$ is related not to  one classical trajectory, but to a family or set of such
trajectories,
$\{\Bq(t)\}$.  
\item  The next order term $U$ 
is related to $S$ by 
\be {2a} \frac{\diff U}{\diff t} + \sum_{k=1}^N\frac{1}{2m_k}\left(\frac{\diff ^2S}{\diff
\Bq^2_k}+ 2\frac{\diff S}{\diff \Bq_k}\Bdot\frac{\diff U}{\diff \Bq_k}\right) =0\;.\ee
Thus $U$, which is the $\hbar$-independent factor in $\Psi, $ is real in the
classically accessible regions, and therefore
 \be {2} w(\Bq, t) \equiv 
\exp[2U(\Bq,t)]=\lim_{\hbar\to 0} |\Psi|^2\;.\ee  When rephrased in terms of
$w$, \eq{2a} becomes a
 classical  continuity equation in $\bC$:
\be {3} \frac{\diff w}{\diff t}+\sum_{k}\frac{\diff }{\diff
\Bq_k}(w\dot{\Bq}_k) =0\;,
\ee where $\dot{\Bq}_k$ is the  velocity at time $t$ of the $k^{{\rm  th}}$ particle
as determined by  the classical equations of motion. Thus $w$ plays the role of a
density and $w\dot{\Bq}$ that of a 3$N$ dimensional current vector in $\bC$; i.e.,  \Eq{3}
is the
$\hbar\to 0$ limit of the Schr\"odinger continuity equation.
\item The quantity $w(\Bq,t)$, apart from an overall arbitrary constant, is  Van Vleck's
determinant $D$, a purely classical quantity (involving derivatives of $S$). For a given
$|\Psi|^2$, $D$ determines how  the trajectories that form the set $\{\Bq(t)\}$  are
``populated.''  

 \end{enumerate}

In short,   in the classical limit the
Schr\"odinger equation does not describe a single system, 
but a population of replicas of such a system moving along the trajectories $\{\Bq(t)\}$.
As $\Psi$ depends only on the degrees of freedom of a single system, and because 
experiments can be done on individual systems,
  this population is to be be visualized as 
   a set of  identical specimens which, one at a time, follow one or another of the
allowed classical trajectories $\{\Bq(t)\}$,  and which, in retrospect,   produced
a population of these trajectories as specified by $w(\Bq,t).$

For any specific solution $\Psi$ of a specific Schr\"odinger equation, that is all that
can be said about the  the trajectories and their population. To have better knowledge of
which trajectories are being followed given this $\Psi$, one must intervene by changing the
Hamiltonian in the Schr\"odinger equation, e.g., with a potential that does not allow 
 trajectories to proceed unless they pass through an aperture of dimension $a$ with edges
smooth enough to leave the WKB approximation valid.\cite{Gutz} Thereafter, the trajectories
and their population will change; i.e., $\Psi$ will change.

The preceding discussion  does not purport to be  a  derivation of the Born
interpretation of
$\Psi(\Bq,t)$. To avert such a misperception, I have  used the word ``population'' instead
of probability. Furthermore, the classical description of what can be extracted from $\Psi$
is only valid as long as the inequality \eq{4} is satified, i.e., if the system is
sufficiently heavy, the forces sufficiently smooth, and the time interval over
which the semiclassical description is needed is sufficiently short. The Born
interpretation has no such restrictions, quite aside from the profound difference between
quantum mechanical probabilities and   probability distributions for a
population of identical systems obeying the laws of classical mechanics.

The restrictions imposed by the semiclassical approximation do not emasculate the
argument being made here.
 For the purpose at hand, it suffices that systems exist that,
on the one hand, have  properties  which allow some of their degrees of freedom  
to be described semiclassically under appropriate circumstances, and on the other,
have inherently nonclassical  degrees of freedom that must be  treated in strict accordance
with the
 laws of quantum mechanics. 

\begin{center}  {\bf III}\end{center}

Consider, then, such a system. The degrees of freedom with a classical counterpart  will
be  the position
$\Bq$ in a configuration space
$\bC$, while the degrees of freedom with no such counterpart live in a finite
dimensional ``internal'' Hilbert space
$\bH$.

 I should explain  that the rules of my game allowed the fictional Newton to know the
{\em  physical phenomena} of electrodynamics when he was asked to unravel the meaning of
the new concepts defined by Maxwell's equations. In the same spirit, the
fictional Maxwell is to know the  phenomena that quantum mechanics supposedly
accounts for. For example, he would be aware of the Stern-Gerlach experiment, and thus know
that if an  atomic species has  a magnetic moment, this moment   only displays a
discrete set of orientations, while from the Schr\"odinger equation he would know that
angular momentum is quantized.

In this scenario, therefore, the internal space of the systems of interest, $\bH$, is
spanned by a finite set of states, which without loss of anything essential can be taken
to be the  spinors 
$\chi_\mu(\Bm)$, where $\Bm$ is the direction of the quantization axis and  $\mu = j, j-1,
\ldots, -j$.  As in the paradigmatic example of the Stern-Gerlach experiment,  the
Hamiltonian  is assumed to contain  a time-independent  perturbation $H_{ {\rm  ext}}(\Bq)$
which   is an operator  both in $\bC$ and $\bH$,
is nonzero only in some bounded region $\bC_{ {\rm  ext}}$ of configuration space,  and is
brought to diagonal form  by  the basis whose quantization axis is 
$\Bn$.  

Let $\psi_{ {\rm  in}}(\Bq,t)$ be a wave function that is accurately approximated by
the WKB Ansatz, and which, for $t<t_1$, describes a set of classical trajectories $\{\Bq_{
{\rm  in}}(t)\}$ that form a beam moving towards $\bC_{ {\rm  ext}}$, which they enter at
$t=t_1.$ Furthermore, let $\chi_\mu(\Bm)$ be the internal state of this beam, so that the
incident  state is represented by
\be  {x1} \Psi_{ {\rm  in}}(\Bq,t) = \psi_{ {\rm  in}}(\Bq,t)\,\chi_{ \mu}(\Bm)\;.   \ee

For $t_1<t<t_2$ this beam traverses the region $\bC_{ {\rm  ext}}$ where the
perturbation $H_{ {\rm  ext}}(\Bq)$ exists, and provided this is sufficiently smooth, the
WKB approximation will continue to be valid. The state \eq{x1} is then  expanded
in  the diagonal basis, and for $t>t_1$ evolves into
\be  {x2} \Psi(\Bq,t) = \sum_{ \sig=-j}^j c_\sig\psi_\sig(\Bq,t)\,\chi_\sig(\Bn)\;, \quad 
 c_\sig =  \left(\chi^*_\sig(\Bn),\chi_\mu(\Bm)\right)\;.
\ee
Each $\psi_\sig$ has a phase which is a solution of  a separate Hamilton-Jacobi equation
with a potential that depends on the quantum number $\sig$, and describes a set of
trajectories
$\{\Bq_\sig(t)\}$ that move in distinct directions. Each such set has a density
$w_\sig(\Bq,t)$ satisfying a separate continuity equation, which means that
\be  {x3} \int d\Bq\; w_\sig(\Bq,t) = C_\sig\;, \qquad t_1<t<t_2\,, \ee
where the constants $C_\sig$ could depend on $\sig$. But  $\psi_\sig(\Bq,t) \to \psi_{
{\rm  in}}(\Bq,t_1)$ as
$t\to t_1$ from above, and therefore all the $C_\sig$ are equal: 
\be  {x4}  \int d\Bq\;w_{ \sig}(\Bq,t)=\int d\Bq\;w_{ {\rm  in}}(\Bq,t) = 1\;,   \ee
where the last equality  is just a  convention.

By design, the region in which the perturbation $H_{ {\rm  ext}}$ acts is large enough
to produce beams that are well separated for $t>t_2$, so that for  such times the
density is \cite{interference}
\be  {x5} w(\Bq,t) = \sum_\sig |c_\sig|^2\,w_\sig(\Bq,t)\;. \ee
Because of \eq{x5}, the fraction of the incident population that ends up in the beam
bearing the label
$\sig$ is 
$|c_\sig|^2\,.$ All but one of these beams, with $\sig=\rho$, can then be eliminated by a
another interaction term in $H$. This filtered  beam can  be sent through a second
arrangement of the Stern-Gerlach variety  with any  other orientation $\Bk$, which will
 produce a set of beams with population fractions 
\be  {x6} |(\chi_\rho^*(\Bn),\chi_\sig(\Bk))|^2\;.  \ee 
All the familiar statements about the complex coefficients $c_\sig$  as probability
amplitudes emerge, therefore, though thus far only expressed as fractions of 
various populations that  pass through some combination of filters and fields.

This then leads to the question of whether a label $``\sig$ along $\Bn$'' can be assigned
to a specimen following a trajectory in the initial set $\{\Bq_{ {\rm  in}}(t)\}$, before
it
 enters the force field that produces the separation into the distinct sets
$\{\Bq_\sig(t)\}$. If this could be done, then each member of the population would have an
{\em inherent} property called
$\sig$, which is {\em revealed} by the subsequent segregation into the distinct sets
$\{\Bq(t)_\sig\}$.  But this cannot be done, because the  intial internal state
$\chi_\sig(\Bn)$ in \eq{x1} can be expanded in any of the infinity of bases in $\bH$, and
the appropriate choice is only revealed {\em  after} the beam has entered the
separating field. \begin{quote}{\em Thus an intrinsic property ``$\sig$ along
$\Bn$'' cannot be assigned to individual specimens; all that can be said is  that if a
specimen passed through the field oriented along
$\Bn$,  the probability that it will emerge in the population ``$\sig$'' is
$|c_\sig|^2.$ }\end{quote}

\noindent This is just the orthodox meaning of probabilities in quantum
mechanics: {\em the probability of a specific outcome as revealed by measurement,}
John Bell notwithstanding.

The same combination of semiclassical and quantum mechanical descriptions can be given for
experiments of the Bohm-EPR type --  for example, a system at rest that disintegrates in
two fragments which then follow opposed classical trajectories $\{\Bq_1(t)\}$ and
$\{\Bq_2(t)\}$, and a suitable correlated state in the joint internal Hilbert space
$\bH_1\otimes\bH_2$. When the widely separated fragments are passed through fields that
produce distinct trajectories for the various eigenvalues $(\sig_1,\sig_2)$ along the
directions $(\Bn_1,\Bn_2),$ they will produce correlations that violate the Bell
inequalities, with all the familiar implications that follow therefrom.

  \vspace{  .5cm} I thank David Mermin for asking several pointed questions.
 
\vspace{  .3cm}


\begin{thebibliography}{99}
\bibitem{Whig} Here I am not engaging in `Whig' history, or any other sort of history, but
considering how we view the accomplishments of the past in the light of present knowledge.
\bibitem{interference} One might worry about interference terms. As soon as the beams cease
to overlap in space, there is no interference. In any event,   there never is any
interference because the internal wave
functions are orthogonal. The  fictional Maxwell would  surely realize this as
he would have known long before about the analagous phenomenon of light  propagation  in
a  medium immersed in a magnetic field.
\end{thebibliography}
\end{document}